\documentclass[12pt,preprint]{aastex}
\usepackage{txfonts}
\usepackage{amssymb}
\usepackage{lscape}

\shorttitle{A Synchrotron Model for 3C 273 Knots} \shortauthors{Liu et al.}

\begin{document}


\title{AN ASYMMETRICAL SYNCHROTRON MODEL FOR KNOTS IN THE 3C 273 JET}


\author{Wen-Po ~Liu\altaffilmark{1,2}, Y. J. ~Chen\altaffilmark{3,4}, and Chun-Cheng ~Wang\altaffilmark{2,5}}


\altaffiltext{1}{Correspondence author; College of Science, Civil Aviation University of China, Tianjin 300300, China; email: wp-liu@cauc.edu.cn}
\altaffiltext{2}{Key Laboratory for Research in Galaxies and Cosmology, Chinese Academy of Sciences, 96 Jinzhai Road, Hefei 230026, Anhui, China}
\altaffiltext{3}{Shanghai Astronomical Observatory, Chinese Academy of Sciences, Shanghai 200030, China}
\altaffiltext{4}{Key Laboratory of Radio Astronomy, Chinese Academy of Sciences, 2 West Beijing Road, Nanjing, Jiangsu 210008, China}
\altaffiltext{5}{Department of Astronomy, University of Science and Technology of China, Hefei, Anhui 230026, China}


\begin{abstract}

To interpret the emission of knots in the 3C 273 jet from radio to X-rays, we propose a synchrotron model in which, owing to the shock compression effect, the injection spectra from a shock into the upstream and downstream emission regions are asymmetric. Our model could well explain the spectral energy distributions of knots in the 3C 273 jet, and  predictions regarding the knots spectra could be tested by future observations.

\end{abstract}



\section{INTRODUCTION}

Many researchers have deeply studied the emission mechanisms of large-scale X-ray jets in active galactic nuclei (AGN) (see the review by Harris \& Krawczynski 2006), but their physical origin is still not resolved. The 3C 273 jet is a nearby (z = 0.158) large-scale jet with a projected length of 57 kpc (e.g., Harris \& Krawczynski 2006) and has been observed in the full band (e.g., Jester et al. 2001, 2005; Uchiyama et al. 2006; Meyer \& Georganopoulos 2014), which is considered a key to unlocking the mystery of large-scale jet emission. Two components of the knots in the 3C 273 jet have been identified by many researchers (e.g., Sambruna et al. 2001; Uchiyama et al. 2006): the emission from radio to optical comes mainly from a synchrotron low-energy component, and the second component including X-rays, could be explained by synchrotron emissions from an additional population of particles or the model of inverse Compton scattering of cosmic microwave background photons (IC/CMB).

Recently, Meyer \& Georganopoulos (2014) showed that the predictions of the IC/CMB model violate the $\emph{Fermi}$ observations, which has ruled out the IC/CMB model for the X-ray component of knot A in the 3C 273 jet. Zhang et al. (2010) pointed out that the IC/CMB model cannot explain the X-rays of knots C1, C2, D1, and D2H3 in the 3C 273 jet. Thus, the X-rays from these knots in the 3C 273 jet are of synchrotron origin. The probable relationship between the two synchrotron components and whether they originate from the same acceleration source or different ones are not yet known. The spectral energy distribution (SED) of the M87 jet from radio to X-rays could be explained by synchrotron emission from a single population of electrons (e.g., Liu \& Shen 2007; Sahayanathan 2008), and the electron spectral index (2.36 on average) agrees well with that predicted by diffusive shock acceleration theory (2.0-2.5; Kirk \& Dendy 2001), which implies that source particles in AGN X-ray jets may originate from a shock. Here, we will build a synchrotron model to interpret the SED (from radio to X-rays) of knots in the 3C 273 jet under the condition of a shock acceleration origin of the source electrons.

In $\S$ 2, we propose an asymmetrical synchrotron model. In $\S$ 3, we apply our model to the SED (from radio to X-rays) of knots in the 3C 273 jet and discuss the model fittings. A conclusion is given in $\S$ 4.

\section{THE MODEL}
In the continuous injection (CI) synchrotron model of Kardashev (1962), a power-law distribution of source particles, which may be generated by a shock, is continuously injected into the emission region with a large-scale equipartition magnetic field $B_{eq}$. As long as the injection interval of source electrons from a shock is shorter than their synchrotron lifetimes in the emission region, the source electrons can be regarded as being quasi-continuously injected into the emission region (Kellermann 1966). Heavens $\&$ Meisenheimer (1987) further considered the advection transport and diffusion of the source electrons and gave a similar spectral shape of the overall emission region as Kardashev (1962), which shows that advection transport and diffusion of the source electrons do not affect the SED of the overall emission region for the CI model. However the CI model cannot explain the two-component emission of knots in the 3C 273 jet from radio to X-rays.

Here, we build a modified CI synchrotron model under the condition of a shock origin of the source electrons: a shock that generates a power-law distribution of source electrons strongly compresses the downstream interface into a very compact region with a stronger magnetic field than $B_{eq}$. (Because of advection transport of the source electrons, the emission region is larger than the shock acceleration region, so its magnetic field is almost unaffected by the shock compression effect and is weaker than that at the downstream interface. Consequently, we assumed that the magnetic field of the emission region is an equipartition magnetic field $B_{eq}$.) The high-energy electrons from a shock would quickly lose energy through a synchrotron process while crossing the very compact downstream interface, which is not resolved by observation. Thus the maximum energy of the injection spectrum into the downstream emission region is lower than that into the upstream one. Considering the irregularity of the downstream magnetic field, the shock compression will lead to a distribution of magnetic field intensities along the cross section of the downstream interface, which results in a broken form of the superimposed injection spectrum into the downstream emission region. For example, we could assume that the magnetic field along the cross section of the downstream interface is $B \propto r^{a}$ (index $a>0$), where $r$ is the distance to the center of the downstream interface. Because of synchrotron loss, the maximum energies of the electrons entering the downstream emission region from different locations on the downstream interface is $E\propto B^{-2}\propto r^{-2a}$ (Kardashev 1962), and the integrated injection spectrum at high energies would be proportional to ($\int E^{-p_1}rdr \propto \int E^{-p_1}dE^{-1/a} \propto E^{-p_2}$)(index $p_2=p_1+\frac{1}{a}>p_1$). However, low-energy electrons entering the downstream emission region from different locations on the downstream interface retain the same spectral shape, i.e., $E^{-p_1}$; thus, the superimposed injection spectrum to the downstream emission region has a broken form.

Thus, the injection spectrum into the upstream emission region is a power law distribution of electrons:

\begin{equation}
Q_u=q_{u}E^{-p_1}, \mbox{~$E_{min}<E<E_{uc}$~,}
\end{equation}

\noindent where, $E_{min}$ and $E_{uc}$ are the minimum and maximum electron energies of the upstream injection spectrum, respectively, and $p_1$ is the electron distributions spectral index.

Because the injection spectrum into the downstream emission region has a broken form, here we consider a broken power law distribution of electrons:

\begin{equation}
Q_d=\left\{
\begin{array}{ll}
q_{d1}E^{-p_1},&\mbox{~$E_{min}<E<E_{db}$~;}\\
q_{d2}E^{-p_2},&\mbox{~$E_{db}<E<E_{dc}$~,}\\
\end{array}
\right. \end{equation}

\noindent where, $E_{dc}$ and $E_{db}$ correspond to the maximum and broken electron energies of the injection spectrum into the downstream emission region, respectively; $p_2$ is the broken spectral index of electrons, and $p_1<p_2$, and $E_{dc}<E_{uc}$.

Here, we are concerned with the SED of an entire knot rather than with the details. We assume that the magnetic field of the emission region is an equipartition field, and the timescale of adiabatic expansion of the source electrons is much longer than their synchrotron lifetimes, so the adiabatic effect can be ignored. Then the kinetic equations of the upstream and downstream emission regions are

\begin{equation}
\frac{\partial N}{\partial t}=\beta\frac{\partial}{\partial
E}(E^2N)+Q_{u,d},
\end{equation}

where $\beta$ is a coefficient representing the synchrotron losses. For simplicity, we assume that the distributions of injection electrons into the upstream and downstream emission regions are isotropic, and the large-scale magnetic field of the emission regions is an equipartition field. We can derive the flux expressions of the emission regions using equation (17) of Kardashev (1962).

For the upstream emission region:

\begin{equation}
I_{u}=\left\{
\begin{array}{ll}
k_{u1}\nu^{-(p_1-1)/2},&\mbox{~$\nu\ll \nu_{ub}$~;}\\
k_{u2}\nu^{-p_1/2},&\mbox{~$\nu_{ub}\ll \nu\ll \nu_{uc}$~,}\\
\end{array}
\right. \end{equation}

$$\nu_{uc}=c_3\beta^2 E^2_{uc}, \nu_{ub}=c_3t^{-2}, c_{3}=3.4\times 10^8 B_{eq}^{-3},$$

\noindent where $t$ (in yr) is the synchrotron lifetime, and $\nu_{ub}$ and $\nu_{uc}$ are the break and maximum frequencies of the upstream emission component, respectively.

For the downstream emission region:

\begin{equation}
I_{d}=\left\{
\begin{array}{ll}
k_{d1}\nu^{-(p_1-1)/2},&\mbox{~$\nu\ll \nu_{db}$~;}\\
k_{d2}\nu^{-(p_2-1)/2},&\mbox{~$\nu_{db}\ll \nu\ll \nu_{dc}$~,}
\end{array}
\right. \end{equation}

$$\nu_{db,dc}=c_3\beta^2 E^2_{db,dc},$$

\noindent where $\nu_{db}$ and $\nu_{dc}$ are the break and maximum frequencies of the downstream emission component, respectively. Here we assume that $\nu_{dc}\ll \nu_{ub}$.

Then, from our equations (4) and (5) and equation (17) of Kardashev (1962), we have
\begin{equation}
\frac{k_{u1}}{k_{d1}}=\frac{q_{u} R_u }{q_{d1} R_d},
\end{equation}

where $R_u$ and $R_d$ are the lengths of the upstream and downstream emission regions along the line of sight, respectively.

If $R_u=R_d$,
\begin{equation}
\frac{k_{u1}}{k_{d1}}=\frac{q_{u}}{q_{d1}}\\.
\end{equation}

\section{FITTING RESULTS AND DISCUSSION}

The data (from radio to X-rays) of knots A, B1, B2/B3, C1, C2, D1, D2/H3, H2, and H1 were taken mainly from Table 1 of Uchiyama et al. (2006), in which a conservative error of 10$\%$ is assigned to most of the data points considering the systematic errors. The data at $1.86\times10^{15}$Hz were taken from Jester et al. (2007), and the X-ray data for knot H2 were taken from Jester et al. (2006). The best fit was obtained using a fitting method similar to that of Liu \& Shen (2007, 2009) and Liu et al. (2013). The data points (except knot H1) from radio to X-rays were arbitrarily divided into four groups, except that we imposed the following requirements: the first group must contain at least radio data based on the SED trend of all the knots, the first two groups must satisfy equation (5) and the first piece of equation (4), and the other two groups must satisfy equation (4). The reduced chi square values, $\chi^2_{\nu}$,  of all the possible combinations were calculated, and the minimum among them is our best fit. In the diffusive shock acceleration theory, the predicted electron spectral index is 2.0-2.5 (Kirk \& Dendy 2001), so we set the range of $p_1$ to be 2.0-2.5. For knot H1, which lacks UV and X-ray data, we fitted its SED by the downstream emission component (i.e., equation (5)).

We plot the data points and model fits in Fig.1, and the possible best fitting parameters are shown in Table 1. Our fitting could well fit the SED (from radio to X-rays) of knots in the 3C 273 jet, which implies the rationality of the kinetic equation (3). The downstream and upstream emission components of the knots are shown by dotted and dashed lines, respectively. The solid line is the sum of the two components, and the vertical line indicates the possible maximum frequency of the downstream emission component. From Table 1, we find that the indices $p_1$ of the inner knots (A, B1, B2/B3) are smaller than those of the outer knots, which may imply that the shock acceleration efficiency of high-energy particles decreases along the jet. According to equation (7) and Table 1, we have ${k_{u1}}/{k_{d1}}={q_{u}}/{q_{d1}}\ll 1$, so source electrons mainly escape along the shock motion direction (i.e., downstream of a knot). The break frequencies $\nu_{ub}$ and $\nu_{db}$ seem to show no regularity along the jet; this shows the complexity of the physical parameter distributions along the jet.

\section{CONCLUSION}
Under the condition of a shock origin of source electrons, we propose a modified CI synchrotron model in which, because of the shock compression effect, the injection spectra from a shock into the upstream and downstream emission regions are asymmetric. Our model fitted the SED (from radio to X-rays) of knots in the 3C 273 jet well, and future observations could test our predictions regarding the spectra of the knots.

\section{ACKNOWLEDGMENT}
The authors thank the anonymous referee for constructive comments that substantially improved an earlier draft. The author (W.-P. Liu) acknowledges the support from the National Natural Science Foundation of China (NSFC) through grant U1231106 and the Open Research Program (14010203) of Key Laboratory for Research in Galaxies and Cosmology, Chinese Academy of Sciences. He also acknowledges the Scientific Research Foundation of the Civil Aviation University of China (09QD15X). The author (C.-C. Wang) acknowledges the support from the NSFC through grant 11421303. This work is also supported by the NSFC (grant 11273042), the Strategic Priority Research Program of the Chinese Academy of Sciences (grant No. XDA04060700) and the Science and Technology Commission of Shanghai Municipality (grant No. 12ZR1436100).

\clearpage

\begin{table}[ht]
\begin{center}
\caption{An Asymmetrical Synchrotron Model Parameters of 3C~273 Jet Knots}
\begin{tabular}{cccccccccc}
%\rotate
\tableline\tableline
Parameter & A & B1 & B2/B3 & C1 & C2 & D1 & D2/H3 & H2 & H1 \\
\tableline
%1 & 2 & 3 & 4 & 5 & 6 & 7 & 8 & 9 & 10 \\
%\sidehead{Low-energy component:}
$\chi_{\nu}^2$ & 2.0 & 0.8 & 2.5  & 0.2  & 1.1   & 1.7   &  3.8  &  2.5 &  0.7   \\
$p_1$ & 2.0 & 2.0  &  2.0  & 2.4  &  2.1  &  2.5  &  2.4  & 2.5  &  2.5   \\
$p_2$ & 3.6 &  4.9 & 4.4  &  5.3   &  4.9  & 4.8  &  4.5  & 6.0  &  ...  \\
$\nu_{db}$ ($10^{12}$Hz) & 1.88 & 8.05  & 5.81  &  40.2   & 13.5  &  26.0  &  8.86  &  13.6  &  9.63   \\
$\nu_{ub}$ ($10^{15}$Hz) & 20.1 & 8.23  & 6.14  &  14.0   & 2.28 &  22.2  & 4.76  &  7.42  &  ...  \\
${k_{u1}}/{k_{d1}}$ ($10^{-3}$) & 27.0 & 8.2  &  15.4 &  70.2   &  8.0  & 26.4  & 14.6  &  3.74 & ... \\
\tableline

\end{tabular}
\end{center}
\end{table}

\begin{figure}
\begin{center}
\includegraphics[scale=0.75] {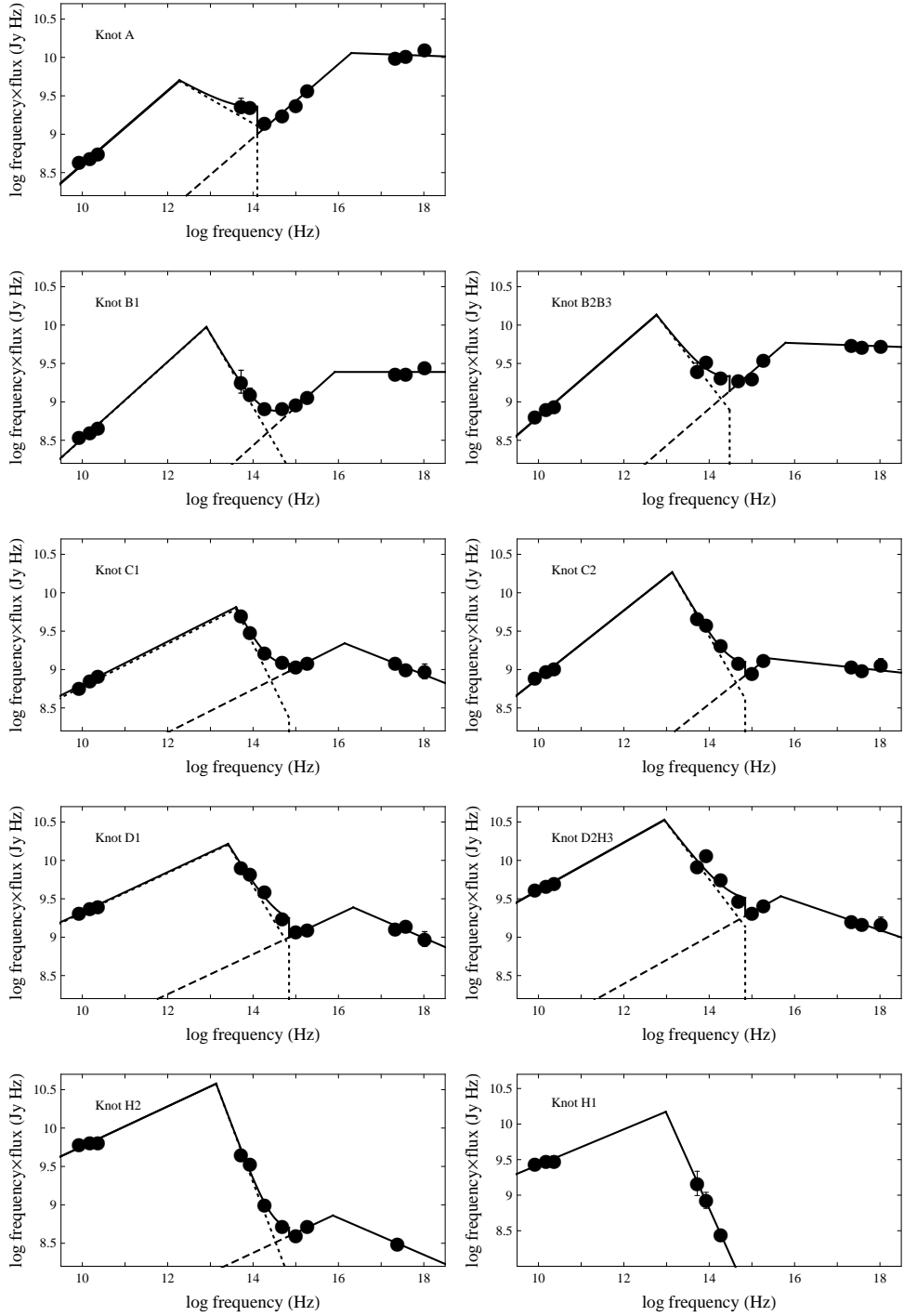}
\caption{SED and model fits of knots in the 3C 273 jet. The horizontal and vertical axes are the logarithms of frequency (Hz) and frequency$\times$flux (Jy Hz), respectively. The data (from radio to X-rays) of knots A, B1, B2/B3, C1, C2, D1, D2/H3, H2, and H1 were taken mainly from Table 1 of Uchiyama et al. (2006). $1.86\times10^{15}$Hz were taken from Jester et al. (2007), and X-ray data for knot H2 were taken from Jester et al. (2006). Dotted and dashed lines show the downstream and upstream emission components of knots, respectively. Solid line is the sum of the two components; vertical line indicates the possible maximum frequency of the downstream emission component.}
\end{center}
\end{figure}
\end{document}